\documentclass[aps, twocolumn, showpacs]{revtex4-1}
\usepackage{amsmath}
\begin{document}
\title{Anomalous magnetic moment and vortex structure of the electron}
\author{S. C. Tiwari \\
Department of Physics, Institute of Science, Banaras Hindu University, Varanasi 221005, and \\ Institute of Natural Philosophy \\
Varanasi India\\}
\begin{abstract}
The electron magnetic moment decomposition calculated in QED is proposed to have origin in a multi-vortex internal structure of the electron.
This proposition is founded on two important contributions of the present work. First, a critical review
on Weyl model of electron, Einstein-Rosen bridge and Wheeler's wormholes leads to a new idea of a topological defect in cylindrical space-time
geometry as a natural representation of the electron. A concrete realization of this idea in terms of the derivation of a new vortex-metric, 
and its physical interpretation to establish the geometric origin of the electron magnetic moment decomposition constitute the second contribution.
Remarkable occurrence of a vortex structure in the basis mode function used in BLFQ is also pointed out indicating wider ramification of 
the proposed electron model.
 
\end{abstract}
\pacs{04.20.-q , 02.40.Ky, 11.15.Tk}
\maketitle

\section{\bf Introduction}

Recent discovery of the Higgs boson at LHC gives strong credence to the Standard Model (SM) of particle physics. The SM has withstood precision 
experimental tests carried out over past decades. The successful development of the gauge theory of electro-weak sector, and the advances in
perturbative QCD methods combined with near absence of new/exotic events at LHC make the quest for unified theories beyond the SM superfluous
from the view-point of the empiricism/operationalism. However the known drawbacks on the conceptual foundations of the SM and the exclusion
of gravity described by general relativity in this unified scheme, i. e. the SM, motivate enormous efforts in superstring theory. Unfortunately superstrings
continue to have tenuous, if any, link with the real particle world. Moreover superstrings seem to have entered into a state of stalemate. Though the 
necessity of a new idea is felt by many physicists the direction of the new pathways is uncertain and speculative: intensely explored ideas
related with, for example, AdS/CFT correspondence and qubits/quantum entanglement. One of the important consequences of these studies is that gauge 
fields/forces as well as space-time may appear as emergent phenomena. The contrasting viewpoints on the nature of space-time 'an arena or it is all' \cite{1}, 
however remain unsettled and open. In this scenario I believe
the profound thoughts of the great minds envisaged in the past could offer a guideline: geometry and topology of the physical space-time have to
be fundamental. We suggest that exploring the internal space-time structure of the electron could serve the basis for a new approach.

Historically classical models of electron were concerned with the extended charge distributions to
explain the electron rest mass, beginning with J J Thomson's model. Lateron spin, Bohr magneton and anomalous magnetic moment also inspired geometric and 
field theoretic models. Electron spin arises naturally in the Dirac equation, and the intrinsic magnetic moment vector ${\bf \mu}$ has the z-component
\begin{equation}
\mu =\pm \frac{e \hbar}{2 m c}=\pm \mu_B
\end{equation}
In terms of g-factor, expression (1) implies that $g=2$ for the Dirac electron. Free electron magnetic moment measured using a one-electron quantum 
cyclotron \cite{2}, however gives a different value
\begin{equation}
\frac{g}{2} =1.00115965218073 (28)
\end{equation}
Compare it with that determined in the landmark experiment \cite{3}
\begin{equation}
\frac{g}{2} =1.001159652200
\end{equation}
Deviation from the Bohr magneton (1) first appeared in 1947 in the hyperfine splitting of energy levels in the Zeeman effect in hydrogen and deuterium 
\cite{4}. Soon after, Schwinger's QED calculation \cite{5} to the first order in the fine structure constant $\alpha =\frac{e^2}{\hbar c}$ gave 
the altered expression
\begin{equation}
\mu_e^\prime =\mu_B[1+\frac{\alpha}{2\pi}]
\end{equation}
Interestingly for the assumed value of $\alpha =\frac{1}{137}$ one gets
\begin{equation}
\frac{g}{2} \approx 1.00116
\end{equation}
Higher order corrections to (4) in QED, and also taking into account the corrections required in the SM \cite{2, 6} show 
extremely precise agreement with the experimental value given by (2).

Let us consider the QED calculated expression for the magnetic moment to the second order in $\alpha$
\begin{equation}
\mu_e =\mu_B [1+\frac{\alpha}{2\pi} - 0.328478444 \frac{\alpha^2}{\pi}]
\end{equation}
Does this structure of the QED expression of the anomalous magnetic moment have deep physical significance? Since the empirical number for
$\mu_e$ (2) gives a measured quantity, from the operational point of view the question of the physical significance of individual terms 
in (4) or (6) would seem to be unimportant or meaningless. However this question has fascinated many physicists, and led to the speculations on 
the internal structure of the electron. Comprehensive review \cite{7} shows that generally it is believed that these attempts failed to offer viable 
alternative to QED and SM. Moreover, high energy electron scattering experiments \cite{8} severely limit the size of the electron to $< ~10^{-16}$cm. 
It could be argued that the failed attempts perhaps indicate that the right conceptual issues were not raised, and new ideas were missing \cite{7}. 
For example, the quantization of charge in the unit of electronic charge $e$ intrigued Dirac to propose magnetic monopole, however the most 
fundamental question as to the nature and origin of charge has not been asked \cite{7, 9}. In a radical departure from the classical models and the QED 
paradigm we proposed a new spatio-temporal bounded field approach. Weyl geometry, topology and deep physical meaning attributed to the form of fine 
structure constant $\alpha$, and the structure of the anomalous magnetic moment expression (4) motivated this approach; see  
the monograph \cite{7}. We have, in a sense, followed the concerns of prominent founders of QED, notably Dirac and Pauli on the foundations of
QED to seek alternatives.

The main ingredients that distinguish our approach from the past ones may be summarized as follows. Massless internal fields with nontrivial topology
constitute the electron, and the rest mass of the electron is not intrinsic but a secondary physical property. Planck constant appears in $\alpha$ with
other fundamental constants $e$ and $c$ in a ratio. Planck constant has the dimension of action but the dimension of 
angular momentum is also the same as that of action. Dimensionless fine structure constant is interpreted as a ratio of angular momenta, 
i. e. $\frac{e^2}{c}$  is interpreted as angular momentum associated with electron. Remarkably, the anomalous magnetic moment (4) could be re-written as
\begin{equation}
\mu_e^\prime =\frac{e}{mc} [\frac{\hbar}{2} + \frac{f}{2}]
\end{equation}
\begin{equation}
f=\frac{e^2}{2 \pi c}
\end{equation}
Logically the bracketted term in (7) should correspond to the angular momentum. We are led to  the hypothesis that electron spin 
has a fractional part $\frac{f}{2}$, if expressed in units of Planck constant, and the electron charge has a mechanical origin as some kind of rotation. 
For further elucidation of this hypothesis see \cite{9, 10}, and vortex model of the spatio-temporal bounded internal fields is developed in \cite{11, 12}.

Physical interpretation of one-particle Dirac equation continues to be of interest primarily at the conceptual level; curiously the significance of 
the classical electron charge radius $r_e =\frac{e^2}{mc^2}$ is least understood or obscure as compared to the Compton wavelength 
$\lambda_c =\frac{\hbar}{mc}$. Even in QED, quite often, the intuitive picture of quantum vacuum effects is related to the size of the Compton wavelength.
It is easily seen that in an alternative outlook \cite{11, 12} both length scales arise in the fine structure constant $\alpha =\frac{r_e}{\lambda_c}$,
and magnetic moment becomes
\begin{equation}
\mu_e^\prime = \frac{e}{2} [\lambda_c +\frac{r_e}{2 \pi}]
\end{equation}

There is a renewed interest in the recent literature \cite{13, 14, 15, 16, 17, 18} on the possible internal structure of electron. 
The present article makes two significant contributions: i) it is suggested that a topological defect in a cylindrical space-time geometry is the most
natural representation of electron, and ii) a new metric, termed vortex-metric, is derived and physically interpreted to explain the geometric
origin of the electron anomalous magnetic moment (7). It is important to note that the metric structure of space-time could be obtained either 
solving a field equation (e. g. Einstein field equation) or by constructing metric independently of any field equation. We adopt the second 
approach in which black holes do not exist. Since Burinskii \cite{13} underlines the role of black holes in elementary particle models 
a brief digression is made to re-interpret the standard Schwarzschild metric in our approach for the sake of clarity.

The paper is organized as follows. In the next section a critical appraisal on the  Weyl model \cite{19, 20, 21}, the Einstein-Rosen bridge 
for a charged particle \cite{22}, and wormholes \cite{1} is made to gain new physical insights on the geometric structure of electron.
It is concluded that a topological defect, cylindrical space-time geometry and fundamental role of angular momentum 
have to be the main ingredients for building electron model. A brief digression is made in Section III to revisit the 
Schwarzschild metric in the present context. A new metric possessing multi-vortex structure is derived in Section IV. Plausible arguments are
presented to relate this metric with the internal space-time structure of the electron. In the last section the proposed new conceptual framework
is put in the perspective of the Kerr-Newman geometric models \cite{13, 18} and recent light-front quantization methods \cite{14, 15, 16, 17, 23, 24}.
In this discussion we are guided by the critique on QED by Dirac \cite{25}, Pauli \cite{26} and Feynman \cite{14}.  The occurrence of vortex-like
structure in the basis light-front quantization (BLFQ) \cite{16} is suggested based on the recent analysis of the Landau problem \cite{27}.
The significance of vortex as a topological defect and Kelvin's vortex atom model \cite{28} show that vortices may serve the basis for explaining elementary
particle spectrum.

\section{\bf Geometric model of electron: charge and mass}

Electric charge, e and mass, m were the only known physical attributes of the electron in the early decades after its discovery in 1897. Understandably 
the physicists during that period, notably Thomson, Poincare, Lorentz and Abraham sought electrodynamical models of the electron \cite{7}. Geometrization
of gravitation in general relativity of Einstein inspired early geometric models of electron. At this point we must emphasize that the geometry in
the modern gauge field theories and SM belongs to the postulated internal spaces different than the physical space-time geometry \cite{29}. In my 
opinion \cite{9} the following fundamental question has not been addressed in the physics literature: What is the physical origin of electric charge?
For example, in QED local U(1) gauge invariance leads to a conserved Noether current, and the Noether charge is a Hermitian operator and the quantum
vacuum is also gauge invariant. This definition amounts to just a formal construct in the hypothetical internal space.

A fresh outlook on the classical models of electron \cite{19, 22} is presented in this section. The first unified theory of gravitation and 
electromagnetism is due to Weyl generalizing the metric Riemann space postulating noncompact homothetic gauge transformation in addition to the general 
coordinate transformations of general relativity. In the monograph \cite{19} Weyl influenced by Mie's theory discusses electron model in Section 32
based on Einstein-Maxwell theory, and in Section 36 based on his unified theory. On the other hand, Einstein and Rosen in 1935 argue that field
singularities are unacceptable, and present an alternative approach discussing electron as well \cite{22}.

Weyl considers the Reissner-Nordstroem solution of the Einstein-Maxwell equations for a static charged sphere
\begin{equation}
ds^2 = - f_{rn} c^2 dt^2 + f_{rn}^{-1} dr^2 + r^2 d\Omega ^2 
\end{equation}
\begin{equation}
f_{rn} = 1-\frac{2 r_g}{r} + \frac{e_g^2}{r^2} 
\end{equation}
\begin{equation}
d\Omega^2 = r^2(d\theta^2 +sin^2 \theta d\phi^2) 
\end{equation}
Here the gravitational radius of the mass m of the electron is
\begin{equation}
r_g =\frac{G m}{c^2} 
\end{equation}
and the gravitational radius of the charge is
\begin{equation}
e_g = \frac{\sqrt{G} e}{c^2} 
\end{equation}
Physical interpretation demands that electron is a real field singularity and the mass of the electron is continuously distributed in the Universe.
However, there is no analog of the electron charge radius, $r_e$.

Weyl also investigates cylindrical geometry and finds that though the electrostatic potential differs from that of the Maxwell theory, the difference is 
appreciable only in the internal space of the electron where the dimension becomes of the order of the gravitational radius of the 
charge $e_g=10^{-33}$cm.

Einstein-Rosen \cite{22} assert that the field singularities are unphysical, and must be avoided as a fundamental principle. Towards this aim
the authors are prepared to modify the field equation. A coordinate transformation is used to obtain regular solutions. For the charged particle
the new metric is obtained such that $f_{rn}$ in (10) is replaced by
\begin{equation}
f_{er} = 1-\frac{2 r_g}{r} - \frac{e_g^2}{2r^2}
\end{equation}
Note the sign change in the last term of (15) as compared to (11) that arises due to the modified field equation. Since mass and charge occur 
independently in the metric tensor, the authors argue that one could set $m=0$.  Thus one may envisage a massless electron model. Assuming
a coordinate transformation replacing $r$ by $u$
\begin{equation}
u^2 =r^2 - \frac{e_g^2}{2} 
\end{equation}
the singularity-free space having two sheets results, and the electron is represented by a bridge between them.

Now returning to the Weyl unified theory we discuss its relatively lesser appreciated part: geometric model of electron discussed in 
Section 36 of \cite{19}. Two markedly distinct features characterize this model. The cosmological constant dependent on the electromagnetic
potentials arises naturally in the field equation. And, the conserved electric current density also depends on the electromagnetic potentials.
In Weyl geometry, a gauge connection $a_\mu$ as a linear 1-form $a_\mu dx^\mu$ gives a distance curvature
\begin{equation}
f_{\mu\nu} =\partial_\mu a_\nu - \partial_\nu a_\mu 
\end{equation}
Weyl assumes $e$ in e.s.u. as a definite unit of electricity interpreting $e a_\mu$ as the electromagnetic potential $A_\mu$. The electric current
density
\begin{equation}
{\bf J} \propto \sqrt{-g} ~ e~ {\bf a} 
\end{equation}
The physical interpretation of electron is speculative and intriguing. Electron is represented by a singularity canal. Charge and mass
characterize singularities and remain constant along the canal. However the world-direction of the canal is determined from the 
Einstein-Lorentz equation, i. e. the geodesic equation generalizing the Newton-Lorentz equation. Weyl dwells on the philosophical arguments
to understand diffused charge throughout the Universe and the sign of electric charge possibly having origin in the arrow of time: past to
future ordering.

To put the import of these classical geometric models in perspective it may be remarked that there exists enormous literature on the singularities
specially once the idea of black hole physics became acceptable among physicists. On the Reissner-Nordstroem solution (10) a nice discussion
given in \cite{30} shows that the singularity at $r=0$ cannot be removed, however other singularities are removable by appropriate coordinate
transformations, and a maximally extended manifold is obtained; see Penrose diagram on page 158 of \cite{30}. Regarding the original Weyl theory
\cite{19} it is known that he himself abandoned it, and entirely different version of gauge symmetry occurs in the modern developments \cite{31}.
Dirac in 1973 revived Weyl geometry for its 'simplicity and beauty' essentially with the motivation of developing his belief in the large number
hypothesis \cite{32}. My own interest in the Weyl-Dirac theory has been to understand the electron structure \cite{7, 9, 10, 21}.

A fresh outlook on the speculations of Weyl \cite{19} and Einstein-Rosen \cite{22} leads to radically new ideas as we discuss below.
First let it be noted that the nature of electric charge as such remains obscure in these works. However new light is thrown on the electron mass.
To understand it we recall an important result in the conventional approach \cite{33} treating electron as a charged sphere and assuming
Reissner-Nordstroem exterior metric. Authors find that the application of the junction conditions on the boundary imply negative mass density
of electron. The argument runs as follows. For large $r$ the third term in the expression (11) is negligible compared to the second term. The value 
of $(\frac{e_g^2}{r^2} - \frac{2 r_g}{r})$ at the assumed boundary $r= 10^{-16}$cm is $\sim 2 \times 10^{-36}$. Thus $f_{rn} > 1$, and the junction
condition shows that the mass density for some value of $r$  in the interior has to be negative. This result puts a question mark on the validity
of the singularity theorems where positivity of mass plays a crucial role \cite{30}. Curiously in the Einstein-Rosen model \cite{21} the 
possibility of massless charge arises as an interesting idea. In Weyl geometry \cite{19} null vectors or gauge-invariant zero length have 
a unique place: the proposition of a massless electron seems quite natural \cite{21}. Thus it may be argued that mass is not an intrinsic 
physical attribute of electron. The origin of observed mass, in that case,  needs a different explanation than what was attempted in 
the past, namely the electrodynamical models.

On the nature of singularities discussed in \cite{19, 22} topological perspective could be another line of thought. Einstein firmly believed that 
singularities in field theories were physically unacceptable see, \cite{34}. He termed the Einstein field equation incomplete, and suggested  
provisional role to the energy-momentum tensor $T_{\mu\nu}$ on the right hand side of the equation
\begin{equation}
G_{\mu\nu} = \frac{8 \pi G}{c^4} T_{\mu\nu} 
\end{equation}
Obviously Einstein-Rosen arguments \cite{22} are in conformity with this belief. A critique \cite{34} on the foundations of the Einstein field
equation (19) shows that geometrization without the field equation is logically admissible. Note that the metric tensor, and
the geodesic equation are sufficient for this purpose. Moreover crucial experimental tests of general relativity are based mainly on the  
Schwarzschild metric \cite{30}. 

The question is: How to get a topological perspective? Nontrivial topology of spacetime finds beautiful expression in the 
concept of wormholes \cite{1}. Authors discuss Reissner-Nordstroem metric and Einstein-Rosen bridge among several topics in this paper.
Wheeler speculates on a model of electron as a collective excitation of the spacetime foam in quantum geometrodynamics \cite{35}. It may,
however be asked whether quantum theory is necessary in a topological approach. Wheeler recognized that this theory \cite{1, 35} was unable to 
explain 'the world of particle physics'. In spite of great advances in topological quantum field theories since then the progress on the issue raised 
by Wheeler, i. e. to make contact with observed elementary particles has remained unsatisfactory. Could it be due to the belief in the 
fundamental role of quantum theory? In a new approach Post abandons
quantum theory altogether and puts forward the concept of quantum cohomology \cite{36}. My own conceptual picture of space-time departs radically
from the past ideas; it is elaborated in a monograph \cite{37}. Here an attempt is made to apply this picture to the problem of electron.

We postulate electron to be a topological defect, namely a vortex, and accord fundamental significance to the angular momentum
$\frac{e^2}{c}$ than the electric charge. Thus a singularity canal in Weyl theory \cite{19} is re-interpreted as a vortex. The second proposition
is also supported from some basic considerations. The most remarkable fact, not commonly recognized, regarding the unit of electricity
$e$ is that it can be factored out from the Maxwell equations. As a consequence the physical description is indistinguishable for 
the Maxwell field $F_{\mu\nu}$ from that of $\frac{F_{\mu\nu}}{e}$. However in the equation of motion of a charged particle factorization
results into an irremovable factor $e^2$ in the Lorentz force expression as it does not cancel. Similarly the expression for the generalized
momentum becomes
\begin{equation}
p_\mu - \frac{e}{c} A_\mu \rightarrow p_\mu - \frac{e^2}{c} a_\mu 
\end{equation}
Note that in the geometric theory of Weyl $a_\mu$ and $f_{\mu\nu}$ have the dimensions of $(length)^{-1}$ and $(length)^{-2}$ respectively. 
Following the conventional approach Weyl multiplies them by $e$ to get the electromagnetic quantities. In contrast to Weyl, in the light of
the preceding arguments we suggest multiplication by $\frac{e^2}{c}$. In general, a suitable angular momentum unit should be used to make 
transformation from geometry to physics.

\section{\bf Schwarzschild metric: new interpretation}

Let us revisit gravitation in this approach. The most studied Schwarzschild metric is spherically symmetric static solution of the vacuum
Einstein field equation. The Schwarzschild metric is given by form (10) replacing $f_{rn}$ by $f_s$
\begin{equation}
f_s = 1 - \frac{2 r_g}{r} 
\end{equation}
Important experimental tests of classical general relativity are based on this metric, and mathematical theory of black holes also evolved from its 
study \cite{30}. The singularity at $r=0$ cannot be removed but the second one at $r=2 r_g$ can be removed by appropriate coordinate transformation,
for example, the advanced or the retarded null coordinates in the Eddington-Finkelstein form of the metric. Two analytic manifolds 
are separated by the surface at $r=2 r_g$. The Kruskal construction and the Penrose diagram are nicely explained in the text \cite{30}.
 
There exist several derivations of the Schwarzschild metric in the literature, that given by Weyl in Section 31 of \cite{19} is an elegant one.
The crucial part in all derivations is common: the identification of the integration constant is based on the correspondence with the Newtonian 
theory of gravitation giving expression (13). Note that the integration constant has a unit of length in the solution of the vacuum Einstein field
equation. Eddington makes a perceptive remark in a footnote on page 87 \cite{38} regarding the objections for using unit of length
for the gravitational mass. Moreover, 
the interpretation of mass as a source in the metric also  does not seem to have a solid foundation \cite{34}. Logically there is no objection 
if we relate the integration constant with the unit of angular momentum. Could we attribute physical significance to the angular moentum?
The suggested role of angular momentum would seem almost heretical due to spherical syymetry and static nature of the metric. 
Nevertheless let us proceed to explore hidden angular momentum in the metric. 

One can construct a quantity having the angular momentum dimension
using purely dimensional arguments to get 
\begin{equation}
L_g = \frac{G \mathcal{M}^2}{c} 
\end{equation}
Defining $L(r) = \mathcal{M} c r$, the expression for $f_s$ becomes
\begin{equation}
f_s = 1- \frac{2 L_g}{L(r)} 
\end{equation}
In this setting the problem of singularity acquires an entirely different version. The angular momentum $L(r)$ can be assigned a limiting minimum
value, let us assume it to be equal to the Planck constant $\hbar$. Thus the singularity at $r=0$ is rendered superfluous. On the other hand,  at
\begin{equation}
L(r) = 2 L_g 
\end{equation}
the singularity does exist. The limit on $L(r)$ being $\hbar$ one gets mass of the order of Planck mass
\begin{equation}
\mathcal{M}^2 = \frac{M_P^2}{2} 
\end{equation}
Recall that in the standard black hole physics instead of the condition (24) one obtains event horizon for any mass defined by the Schwarzschild
radius
\begin{equation}
r_S = \frac{2 G M}{c^2} 
\end{equation}
whereas the singularity defined by (25) occurs at a definite Planck scale $\sim 10^{19}$GeV. In this re-interpretation the problem of
negative mass density \cite{33} is rendered a nonissue.

In Cosmological models the incorporation of Mach principle has attracted attention of some physicists; Brans and Dicke \cite{39} discuss in 
detail different versions of this principle, and relate it with the following 
\begin{equation}
\frac{ G M_{Universe} }{R c^2} \sim 1
\end{equation}
Here $M_{Universe}$ is the finite mass of the visible Universe and $R$ is its radius. Authors provide a simple argument for the derivation of (27).
Newton's theory gives acceleration due to gravity at a distance $r$ from, let us say the Sun, $\frac{G M_{Sun}}{r^2}$, and from 
dimensional considerations one can construct acceleration to be $\frac{M_{Sun} R c^2}{M_{Universe} r^2}$; equating the two the relation (27) 
follows immediately. In the angular momentum approach, the relation (27) is interpreted as the upper limit on $L(r)$ leading to the maximum value
\begin{equation}
L(r)_{max} =\frac{ G M_{Universe}^2}{c} 
\end{equation}
The ratio between the maximum and minimum values of $L(r)$ turns out to be a large number $\sim 10^{120}$ assuming the rough estimates of the mass and 
the radius of the visible Universe. Further discussion on the large scale structure of the Universe is deferred to a separate paper since the
focus here is on the electron structure.

\section{\bf Electron structure: geometry and topology}

Let us briefly mention the salient features of quantum geometrodynamics \cite{35}. Wheeler's qualitative picture of the gravitational
field fluctuations is in analogy to QED vacuum fluctuations. Feynman's path integral approach is adopted in which the action integral
in the phase exponent uses the Planck constant as a quantum of action following the standard QFT practice. Virtual particle pairs having
charge $\sim 12 e$ and mass of the order of Planck mass are expected to be created by vacuum fluctuations. However such particles cannot be 
identified with the electron or any observed elementary particle in nature. However drawing analogy with QED renormalization method ingenious
arguments are put forward by Wheeler to re-interpret the virtual particles in the wormhole picture.

Departing from Wheeler's approach we propose to incorporate topology of space-time at a fundamental level. For this purpose, 
quantum theory/QED paradigm is shifted to the old quantum theory. To appreciate the significance of the old quantum rule of Bohr
on angular momentum quantization we refer to a short historical account given in \cite{40}, and for its relationship with topology to the 
Post's monograph \cite{36}. Bohr-Wilson-Sommerfeld (BWS) quantization requires nontrivial topology; Post highlights an important contribution 
of Einstein in this connection \cite{36, 41}. Einstein replaces Bohr's circular orbits by torus, and the orbital manifold (orbifold) signifies
topological obstruction. Orbifolds have de Rham cohomology ramifications \cite{36, 41, 42}: the principal result of this work is that 
physical BWS quantum conditions on angular momentum have natural mathematical counterpart in the form of de Rham periods for topological
defects. Application of this idea can be found in the monograph \cite{36} and for topological photon in \cite{43}. Thus the fundamental
significance accorded to Planck constant as a quantum of angular momentum becomes logically justified.

Admittedly introducing $\hbar$ in the Schwarzschild metric is heuristic, however it can be pursued further qualitatively for other metrics.
Reissner-Nordstroem, and Einstein-Rosen metrics become 
\begin{equation}
f_{rn} = 1 - \frac{2 L_g^e}{\hbar}[ 1- \frac{\alpha}{2}] 
\end{equation}
\begin{equation}
f_{er} = 1 - \frac{2 L_g^e}{\hbar}[ 1+ \frac{\alpha}{4}]
\end{equation}
where we define 
\begin{equation}
L_g^e = \frac{G m^2}{c} 
\end{equation}
Apart from the difference in the numerical factor, i. e. 4 instead of $2 \pi$ formal resemblence of the bracketted term in (30) with the 
anomalous part in the magnetic moment expression (4) seems curious. Does it have some deep significance? Let us explore this question.
The numerical value of $L_g^e$ is a very small number; the ratio $\frac{L_g^e}{\hbar}$ is of the order of $10^{-44}$. Note that the electron 
charge radius and the Compton wavelength can be defined using angular momenta $\frac{e^2}{c} =m c r_e$ and $\hbar = mc \lambda_c$ respectively.
A length analogous to them using $L_g^e$ can be defined
\begin{equation}
\lambda_G = \frac{L_g^e}{mc} 
\end{equation}
Since $r_e = \alpha \lambda_c$ we anticipate
\begin{equation}
\lambda_G = \alpha^N ~ \lambda_c 
\end{equation}
for some integer N; in numbers $\lambda_c \sim 10^{-11}$ cm and $\lambda_G \sim  10^{-55}$cm. Following geometric picture presents itself: a core
sphere of radius $\lambda_c$ surrounded by concentric thin spherical shells of width $\alpha^n \lambda_c ; n= 1, 2,....N$.

The concentric spherical shell model shows that multiple geometric sub-structures have to be explored for the electron model: a possible 
propagating vortex is suggested for this purpose. The crucial question is: How do we get it in the metric tensor for the geometric model?
It is obvious that the cylindrical metric rather than the spherical Reissner-Nordstroem metric has to be considered. Unfortunately the known 
Weyl metric \cite{19} did not succeed as electron model. Moreover, the problem of logarithmic divergence at large radial distance for a 
cylindrical system makes physical interpretation very difficult. For the envisaged vortex we adopt an alternative method proposed in \cite{34}.
In this approach massless scalar field $\Phi$  satisfies wave equation in flat spacetime with the metric $\eta_{\mu\nu}$
\begin{equation}
\partial_\mu \partial^\mu \Phi =0 
\end{equation}
and assumed to consist of two scalars
\begin{equation}
\Phi = F \Psi 
\end{equation}
such that $\partial_\mu \Psi$ is a null vector
\begin{equation}
\partial_\mu \Psi\partial^\mu \Psi =0
\end{equation}
The choice for a massless scalar field as a fundamental field variable is primarily based on the guiding principles of simplicity and symmetry.
The proposition (35) gives a bi-scalar structure to the field $\Phi$ necessary to construct a nontrivial metric.
Now the metric tensor is defined in the Kerr-Schild form
\begin{equation}
g_{\mu\nu} =\eta_{\mu\nu} + F \partial_\mu \Psi\partial_\nu \Psi 
\end{equation}
Note that $\partial_\mu \Psi$ is also a null vector with respect to $g_{\mu\nu}$. Earlier we have derived some known metrics, for example, Vaidya metric, 
the Schwarzschild metric, and Brinkman-Robinson metric using this approach \cite{34}.

New solutions in cylindrical system $(\rho, \phi, z)$ with the aim of realizing vortex are presented here. Assuming $\Psi$ to be a function of 
(t, z) only the null condition is satisfied for functions $\Psi(z \pm ct)$. We assume simplest form for a propagating field
\begin{equation}
\Psi = e^{i(k z - \omega t)} 
\end{equation}
Assumed time-independent $F(\rho, \phi, z)$ the wave equation (34) becomes
\begin{equation}
\frac{\partial^2 F}{\partial \rho^2} +\frac{1}{\rho} \frac{\partial F}{\partial \rho} + \frac{1}{\rho^2} \frac{\partial^2 F}{\partial \phi^2}
+ 2 i k \frac{\partial F}{\partial z} +\frac{\partial^2 F}{\partial z^2} =0
\end{equation}
The usual azimuthal-dependence $e^{i l \phi}$ reduces Eq.(39) to
\begin{equation}
\frac{\partial^2 F}{\partial \rho^2} +\frac{1}{\rho} \frac{\partial F}{\partial \rho} - \frac{l^2}{\rho^2} F
+ 2 i k \frac{\partial F}{\partial z} +\frac{\partial^2 F}{\partial z^2} =0
\end{equation}
To solve Eq.(40) we make a simplification that F is not z-dependent. In this case the solution is given by
\begin{equation}
F = \rho^l ~ e^{i l \phi} 
\end{equation}
and the line element assumes the form
\begin{equation}
ds^2 =- c^2 dt^2 + d\rho^2 +\rho^2 d\phi^2 +dz^2 + F(\rho, \phi) (\frac{\partial \Psi}{\partial z})^2 (c dt - dz)^2  
\end{equation}
The solution (41) does represent a vortex: phase singularity at $\rho = 0$. Unfortunately the divergence as $\rho \rightarrow \infty$
is worse than the logarithmic divergence making this metric physically unacceptable.

Retaining z-dependence of $F$ in Eq.(40), and in analogy to the paraxial approximation in optics \cite{44} neglecting $\frac{\partial^2F}{\partial z^2}$
compared to $\frac{\partial F}{\partial z}$ Eq.(40) becomes
\begin{equation}
 \frac{\partial^2 F}{\partial \rho^2} +\frac{1}{\rho} \frac{\partial F}{\partial \rho} - \frac{l^2}{\rho^2} F
+ 2 i k \frac{\partial F}{\partial z} =0
\end{equation}
This equation has the well-known Laguerre-Gaussian (LG) mode solution
\begin{equation}
F(\rho, \phi, z) = LG^l_p 
\end{equation}
Here azimuthal index $l$, and the index $p$ of the associated Laguerre polynomial characterize the vortex and the concentric rings in the 
transverse profile of the LG mode. Two of the remarkable properties of LG modes noted in optics literature \cite{44} are the z-dependent radius
of the beam, and the topological quantum number $l$. Phase singularity has a topological charge $\frac{1}{2 \pi} \oint l d\phi $. In the 
present case, the ansatz (37) provides a geometric significance for the LG mode function in a new line element
\begin{equation}
 ds^2 =- c^2 dt^2 + d\rho^2 +\rho^2 d\phi^2 +dz^2  +LG^l_p (\frac{\partial \Psi}{\partial z})^2 ( c dt - dz)^2
\end{equation}
Note that the metric is not complex; it is understood that following the convenient practice only complex representation is used for 
the solution of the wave equation.

Could we visualize internal structure of the electron geometrically using this metric? Qualitative interesting correspondence follows immediately.
The metric for $l=1, p=0$ shows vortex structure of the form $\rho L^1_0 (\rho^2) ~ e^{-\frac{\rho^2}{w^2}} e^{i \phi}$. Tentative
identification of $w \rightarrow \lambda_c$ makes this vortex as a core vortex for the electron. The concentric ring for $p=1$ may be 
identified with the second vortex having dimension equal to $r_e$. This picture nicely fits the two-vortex Dirac electron model \cite{12}.
The multi-vortex structure can be envisaged for the geometry of the metric (45). To associate angular momentum with the vortex we cannot adopt
the optics method: orbital angular momentum per photon $l \hbar$ is obtained using Poynting vector for linear momentum density and calculating
the orbital angular momentum from it \cite{44}. In the geometric formulation we have to seek a new idea. The geometric quantities like the 
curvature tensors can be calculated following lengthy but straightforward technical method \cite{38}. However these are of no interest for angular momentum.
In view of the structure of the Kerr-Schild form (37) a new geometric quantity can be naturally defined
\begin{equation}
v_\mu = F \partial_\mu \Psi 
\end{equation}
\begin{equation}
L_{\mu\nu} = \partial_\mu v_\nu - \partial_\nu v_\mu = \partial\mu F \partial_\nu \Psi - \partial_\nu F \partial_\mu \Psi
\end{equation}
The antisymmetric tensor $L_{\mu\nu}$ has formal similarity with the electromagnetic field tensor for a specific choice of $A_\mu$ having the
form (46). To get its physical interpretation it is instructive to make explicit calculation of its components: $L_{0i}$ and 
$ L_{ij}; i, j =1, 2, 3$ are nonvashing. Of particular interest is the component $L_{23}$
\begin{equation}
L_{23} = - l k F \Psi 
\end{equation}
Here we take $\partial_2 = \frac{\partial}{\partial \phi}$. We also calculate the nonvanishing components of $v_\mu$
\begin{equation}
v_0 =- \frac{i \omega}{c} F \Psi 
\end{equation}
\begin{equation}
v_3 = i k F \Psi 
\end{equation}
An interesting quantity is the following ratio
\begin{equation}
|L_{23}/v_0| = |L_{23}/v_3| = l 
\end{equation}
Multiplication by appropriate angular momentum unit in (51) provides physical interpretation of this ratio as qunatized angular momentum. 
Note that the ratio $|L_{02}/v_0| =l$. In our approach we have two fundamental units $\hbar$ and $\frac{e^2}{c}$, therefore, the vorticity of
of two different vortices $p=0,~ l=1$ and $p=1, ~ l=1$ are suggested to be quantized in these units respectively. A qualitative 
geometric model of electron inspired by the QED calculated electron magnetic moment written in the form (7) is thus established.

\section{\bf Discussion and Conclusion}

Geometric models of electron have attracted attention of many physicists in the past hundred years: their present status remains tentaive and
speculative. In spite of this there are at least two strong reasons to continue efforts in this direction. First, geometry and topology as 
foundations of physics has tremendous philosophical appeal. Unfortunately there is no breakthrough in superstrings and M-theory in this 
direction; there is a need for radical alternative. Secondly the great advances
in modern quantum field theories have so far failed to address the conceptual issues raised by eminent founders of QED, e. g. Dirac \cite{25}
and Pauli \cite{26}. Candid remarks in the concluding part of Pauli's Nobel Lecture \cite{26} are worth reproducing in this connection: 
``At the end of this lecture I may express my critical opinion, that a correct theory should neither lead to infinite zero-point energies
nor to infinite zero charges, that it should not use mathematical tricks to subtract infinities or singularities nor should it invent a 
'hypothetical world' which is only a mathematical fiction before it is able to formulate the correct interpretation of the actual world of physics.
From the point of view of logic, my report on 'exclusion principle and quantum mechanics' has no conclusion. I believe that it wil only
be possible to write the conclusion if a theory will be established which will determine the value of the fine-structure constant and will
thus explain the atomistic structure of electricity, which is such an essential quality of all atomic sources of electric fields actually occurring
in Nature.''

The criticisms of Dirac and Pauli, in my view, motivate us to discover underlying elements of reality in QED rather than its rejection altogether.
Alternative approach based on geometry and topology has a potential to address fundamental questions. One example that we explore in this paper
is the geometric interpretation of the electron magnetic moment: it is related with the question raised by Feynman at the 1961 Solvay Conference
highlighted recently \cite{14, 15}. Feynman's question is: do individual terms in the perturbative QED calculations of the magnetic moment have
intuitive physical interpretation? A vortex in the multi-vortex structure is proposed to correspond to the individual terms in the electron 
magnetic moment expression. The utility of the present ideas lies in offering a framework for a synthesis of geometry-based approach and QED.
Burinskii in a series of papers, see references in \cite{13}, has attempted developing the Kerr-Newman (KN) geometric model of electron \cite{18}.
A rotating bubble model as an oblate ellipsoid of dimension $\lambda_c$ and thickness $r_e$ is obtained. The author suggests connection between 
black holes and elementary particles, and asserts that extended electron model must be taken seriously. Aspects of superstrings and SM are also
viewed in the light of electron model. Though some interesting points emerge from Burinskii's contributions, it is fair to state that his electron 
model remains tentative. Further progress in his approach may need radically different outlook: incorporating vortex-metric and the angular 
momentum framework. In fact, the Kerr-Schild form of the KN metric, and emergence of a vortex in his analysis are attractive features in this 
respect. Regarding black holes the angular momentum perspective proposed in Section III substantially alters its physics, and more closely
relates it with particle physics in view of topological quantization a la BWS quantization of orbifolds.

Nonperturbative QFT may get viable framework based on LFQ methods \cite{23, 24}. In the present context, the calculation of electron
magnetic moment using LFQ methods hints at internal structure of the electron \cite{14, 15, 16}. A novel development is the basis LFQ (BLFQ)
nonperturbative approach aimed at QCD; its application to QED  \cite{16} has led to a significant result calculating expression (4). This result
signifies 'a nontrivial structure of the electron in QED' \cite{16}. Authors represent a physical electron by a truncated Fock-sector expansion
\begin{equation}
|e_{phys}> = a |e> +b |e \gamma> 
\end{equation}
Two-dimensional (2D) harmonic oscillator wavefunction for the transverse direction and a plane wave for the longitudinal direction
serve as the basis mode function for a Fock particle. Here
\begin{equation}
\Phi_{nm}(\rho, \phi) = C e^{i m \phi} \rho^{|m|} e^{-\frac{\rho^2}{2}} L_n^{|m|} (\rho^2) 
\end{equation}
The mode function (53) has interesting properties: the Fourier transform of coordinate and momentum space wavefunctions have the
same structure, it has finite transverse size, and it has vortex structure \cite{27}. The first two properties have been noted in the
BLFQ literature \cite{16}, however the last one has not received attention. Its significance has recently been pointed out in \cite{27}.
The choice of Cartesian coordinate system results into 2D harmonic oscillator wavefunction as a product of Hermite-Gaussian (HG) functions:
absence of vortex and ill-defined or zero angular momentum characterize this solution. Now, Zhao et al \cite{16} explain that
the mode function (53) as such is not obtained from the field equation, however the plausibility arguments justify 2D oscillator
wavefunction. Obviously the BLFQ calculation needs to be carried out for HG modes and compared with that of (53). We suggest the role of 
vortex in BLFQ deserves investigation. It may be pointed out that the paraxial LG mode function (44) differs from (53) due to its z-dependence. 

The idea that electric charge has mechanical origin in the second part of the expression (7) \cite{7} is given a sound basis in the 
present work. Its physical consequence in the context of Wheeler's quantum geometrodynamics \cite{35} is illuminating. Qualitative 
considerations on the quantum fluctuation of field and integrated flux through wormholes leads to the electric charge of the order of 
$\sqrt{\hbar c} \sim 12 e$ independent of the wormhole size. Since this value of charge looks unphysical Wheeler seeks interpretation
invoking quantum vacuum effects of QED. In contrast, the present geometric interpretation leads to a charge for the core vortex with spin
$\frac{\hbar}{2}$ of value $g= \sqrt{\frac{137}{2}} ~e$. One may calculate charge radius for $g$ to be $\frac{\lambda_c}{2}$. In 
our geometric picture of the internal structure of the electron both charges represent internal vortices. Regarding electric charge 
quantization, specific topological quantum number $l=1$ for the vortex and N as algebraic sum of N vortices gives $Ne$.

An important question is to explain half-integral electron spin since $l$ is an integer. In the KN geometric model Lopez \cite{18} simply
assumes empirically given value of $\frac{\hbar}{2}$. It could be done here also, however in the vortex model it must have origin
in the nature of the topological obstruction. There is no definite result, however for a clue  we suggest insights based on the topological
approach to the proton spin puzzle \cite{45}. Moreover, the vortex structure in the 2D simple harmonic oscillator recently pointed out 
\cite{27} and the occurrence of half-integral representation of the rotation group in 2D oscillator \cite{46} offer a possible link
to resolve this question.

Another issue not addressed in this paper is concerned with the explanation of electron mass. We have argued that mass is a secondary 
physical attribute of electron, not an intrinsic one. A tentaive suggestion is that it may be explained in terms of
vortex-vortex interaction resulting into some kind of spiraling vortex characterized by mass parameter. The problem is being further
investigated.

In conclusion, the peculiar combination of the fundamental constants $(e, \hbar, c)$ in $\alpha$ and the structure of the QED electron 
magnetic moment decomposition motivate us to propose internal space-time vortex structure of the electron. A new vortex-metric and delineation 
of vortex in BLFQ are some of the important contributions of the present work. The conceptual framework proposed here may have significant
implications on the recent developments \cite{ 13, 14, 15, 16, 17}. In particular we emphasize that adopting BWS quantum conditions 
has two important advantages: counter-intuitive quantum weirdness does not exist,  and it is more suitable for a topological approach \cite{47}.
Knotted structures of the fundamental electron vortex could be envisaged
to account for higher spin elementary particle spectrum reviving aspects of Wheeler's wormholes \cite{1, 35} and Kelvin vortex model \cite{28} in the 
light of the present framework in a unifying picture.

{\bf Acknowledgment}

I thank the referees for useful comments that helped improve the presentation of the manuscript.

\end{document}